\documentclass[twocolumn,12pt]{iopart}

\usepackage{graphicx}
\begin{document}

\title[]{Hyperconcentration for multipartite entanglement via linear optics}

\author{Xihan Li$^{1,2}$ and Shohini Ghose$^{1,3}$}

\address{$^1$Department of Physics and Computer Science, Wilfrid Laurier University, Waterloo, Canada N2L 3C5\\
$^2$ Department of Physics, Chongqing University,
Chongqing, China 400044 \\
$^3$ Institute for Quantum Computing, University of Waterloo, Canada N2L 3G1}
\eads{\mailto{xihanlicqu@gmail.com}, \mailto{sghose@wlu.ca}}
\vspace{10pt}

\begin{abstract}
We present a hyperconcentration scheme for nonlocal $N$-photon hyperentangled Greenberger-Horne-Zeilinger states. The maximally hyperentangled state, in which $N$ particles are entangled simultaneously in the polarization and the spatial mode, can be obtained with a certain probability from two partially hyperentangled states. The hyperconcentration scheme is based on one polarization parity check measurement, one spatial mode parity check measurement and $N-2$ single-photon two-qubit measurements. The concentration only requires linear optical elements, which makes it feasible and practical with current technology.
\end{abstract}

\section{Introduction}
Entanglement is an important resource in quantum information processing, and is widely used in quantum communication and computation, including quantum key distribution\cite{qkd1,qkd2,qkd3}, quantum secret sharing \cite{book}, quantum dense coding \cite{dense}, quantum teleportation \cite{tele}, quantum secure direct communication \cite{qsdc1,qsdc2,qsdc3}, quantum repeater \cite{repeater} and so on. Entanglement can be generated in different degrees of freedom (DOF) of physical entities such as photons, electrons, atoms, etc. Among these, the photon is an interesting candidate for quantum communication due to its manipulability and high-speed transmission. The photon has many DOFs to carry quantum information, such as spatial modes, time-bins, polarization, frequency, and orbital angular momentum. Besides the conventional entanglement in which photons are entangled in a single DOF, there is the possibility of hyperentanglement in which photons are entangled in more than one DOF \cite{hyper1,hyper2,hyper3}. Hyperentangled states can increase the capacity of quantum information processing since each photon carries more than one qubit \cite{dense1,qkd,computation,hyper4, hyper5, hyper6}. Moreover, hyperentanglement has some important applications in entanglement purification \cite{puri pdc,puri1,depp,odepp,odepp1,omdepp} and state analysis \cite{bsa1,bsa2,bsa3,bsa4,bsa6,gsa}.

In most quantum communication schemes based on entanglement, maximal entanglement is required to ensure efficiency and security.
However, maximal entanglement is fragile and in practice, it is difficult to preserve during transmission and storage. The inevitable interaction with the environment degrades the fidelity and degree of entanglement of the quantum state, which subsequently affects the security and efficiency of quantum communication protocols. Many solutions have been proposed to recover high quality entangled states from polluted less-entangled samples. One example is entanglement concentration, which extracts a maximally entangled state from an ensemble of less-entangled pure states. The first entanglement concentration scheme was proposed in 1996 based on the Schmidt projection method \cite{concen1}. Later, two entanglement concentration schemes based on entanglement swapping were proposed \cite{concen swap1,concen swap2}. In 2001, entanglement concentration schemes using linear optical elements were proposed and demonstrated \cite{concen pbs1, concen pbs2}. During the past few years, many interesting entanglement concentration schemes considering different physical systems, different entangled states and exploiting different components have been discussed in the literature \cite{concen sheng1,concen sheng4,concen deng,concen sheng2,concen w,concen w5,concen w6}. These entanglement concentration schemes can be classified into two groups based on whether the parameters of the less-entangled states are known or not. If the parameters are unknown, an ensemble of less-entangled states is required to distill maximal entanglement. Otherwise, additional states can be prepared or optical elements can be manipulated according to the known parameters, to accomplish the concentration.

Since hyperentanglement has increasing applications in quantum information processing, the concentration of hyperentangled states has attracted much attention recently. In 2013, Ren \emph{et al.} proposed two hyperentanglement concentration schemes for a two-photon four-qubit system, in which only linear optics was required \cite{hc1}. They also proposed a hyperentanglement concentration scheme assisted by diamond NV centers inside photonic crystal cavities \cite{hc11}. One of us also proposed two hyperconcentration schemes with known and unknown parameters, respectively \cite{hc2}. 
Recently, a general hyperentanglement concentration was also proposed \cite{hc4}. 

In this paper, we present the first hyperconcentration scheme for $N$-photon hyperentangled Greenberger-Horne-Zeilinger (GHZ) states, which are simultaneously entangled in both the polarization and spatial mode DOFs. The scheme uses two less-entangled states and the concentration is realized with linear optics elements only. After one polarization parity check measurement, one spatial mode parity check measurement and $N-2$ single-photon measurements, the $N$ parties share the maximally hyperentangled GHZ states with a certain probability. The scheme can be implemented across an $N$-party network, where the parties are remotely located and do not need to interact with each other.  A notable point is that the hyperconcentration success probability does not decrease with the number of photons - it remains the same as the two-photon success probability. We also discuss possible sources of error and how to address them in a practical setting. 

\section{Parity check devices for different degrees of freedom}
Before we demonstrate our hyperconcentration scheme, we introduce two parity check devices for the polarization and the spatial mode DOFs, respectively. These parity check devices are used to select the even-parity states in the given DOF and then measure one photon both in the polarization and the spatial mode DOFs.
\begin{figure}[!h]
\includegraphics[height=2in]{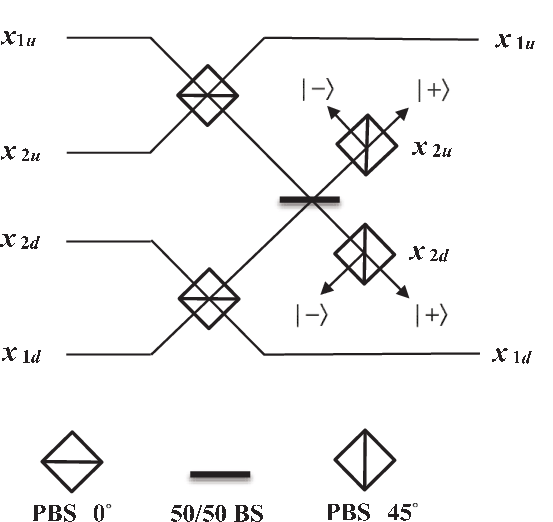}
\caption{The schematic diagram of the polarization parity check device. The photon $X_1$ ($X_2$) enters the apparatus via two potential spatial modes $x_{1u}$ and $x_{1d}$ ($x_{2u}$ and $x_{2d}$). The PBS oriented at $0^o$ transmits the $V$ state and reflects the $H$ state. Then the photon in $x_{2u}$ or $x_{2d}$ is projected onto the diagonal basis of spatial modes by the beam splitter and then measured in the polarization diagonal basis $\vert \pm\rangle$ with the help of PBS at $45^o$, which transmits the $\vert +\rangle$ and reflects $\vert -\rangle$ state. }
\end{figure}
A schematic for the polarization parity check (PPC) is shown in Fig.1.
We start with four possible spatial modes for two input photons $1u, 1d, 2u$ and $2d$.
We use two different kinds of polarizing beam splitters (PBS). The PBS oriented at $0^o$ transmits the vertical polarization state $\vert V \rangle $ and reflects the horizontal one $\vert H \rangle$. The PBS at $45^o$ transmits the $\vert +\rangle=\frac{1}{\sqrt{2}}(\vert H\rangle+\vert V\rangle)$ and reflects the $\vert -\rangle=\frac{1}{\sqrt{2}}(\vert H\rangle-\vert V\rangle)$. And the effect of the 50:50 beam splitter (BS) can be described as
\begin{eqnarray}
In_u\rightarrow \frac{1}{\sqrt{2}}(Out_u+Out_d), \\
In_d\rightarrow \frac{1}{\sqrt{2}}(Out_u-Out_d).
\end{eqnarray}
Here $In_u$ and $In_d$ are the up and down input ports, while $Out_u$ and $Out_d$ are the two output ports of the BS.
The PBS at $0^o$ is used to compare the polarization parity of the two input photons. The BS is used to measure the spatial modes in the diagonal basis $\vert \pm'\rangle=\frac{1}{\sqrt{2}}(\vert u\rangle\pm \vert d\rangle)$ while the PBSs at $45^o$ are utilized to measure the polarization in the diagonal basis $\vert \pm\rangle$. The evolution of different possible input states are
\begin{eqnarray}
&&\vert HH\rangle\otimes (\vert x_{1u}\rangle+\vert x_{1d}\rangle)(\vert x_{2u}\rangle+\vert x_{2d}\rangle)\nonumber\\
&\rightarrow& \vert HH\rangle\otimes (\vert x_{1u}x_{2u}\rangle+\vert x_{1u}x_{2d}\rangle+\vert x_{1d}x_{2u}\rangle+\vert x_{1d}x_{2d}\rangle),\\
&&\vert HV\rangle\otimes (\vert x_{1u}\rangle+\vert x_{1d}\rangle)(\vert x_{2u}\rangle+\vert x_{2d}\rangle)\nonumber\\
&\rightarrow& \vert HV\rangle\otimes (\vert x_{1u}x_{1u}\rangle+\vert x_{1u}x_{1d}\rangle+\vert x_{1d}x_{1u}\rangle+\vert x_{1d}x_{1d}\rangle), \\
&&\vert VH\rangle\otimes (\vert x_{1u}\rangle+\vert x_{1d}\rangle)(\vert x_{2u}\rangle+\vert x_{2d}\rangle)\nonumber\\
&\rightarrow& \vert VH\rangle\otimes (\vert x_{2u}x_{2u}\rangle+\vert x_{2u}x_{2d}\rangle+\vert x_{2d}x_{2u}\rangle+\vert x_{2d}x_{2d}\rangle), \\
&&\vert VV\rangle\otimes (\vert x_{1u}\rangle+\vert x_{1d}\rangle)(\vert x_{2u}\rangle+\vert x_{2d}\rangle)\nonumber\\
&\rightarrow& \vert VV\rangle\otimes (\vert x_{2u}x_{1u}\rangle+\vert x_{2u}x_{1d}\rangle+\vert x_{2d}x_{1u}\rangle+\vert x_{2d}x_{1d}\rangle).
\end{eqnarray}

From the above expression, we find that for the odd-parity states $\vert HV\rangle$ or $\vert VH\rangle$, there are zero or two photons detected by these four detectors set up on paths $x_{2u}$ and $x_{2d}$. However, for the even-parity polarization states $\vert HH\rangle$ or $\vert VV\rangle$, there is one and only one photon detected by these detectors in principle. Therefore, we can distinguish the parity of polarization states according to the detection in the output ports.
With the help of a BS and the PBS oriented at $45^o$, the photon appearing at the output port 2 can be measured in the diagonal basis in both the polarization DOF and the spatial one.

\begin{center}
\begin{figure}[!h]
\includegraphics[height=2in]{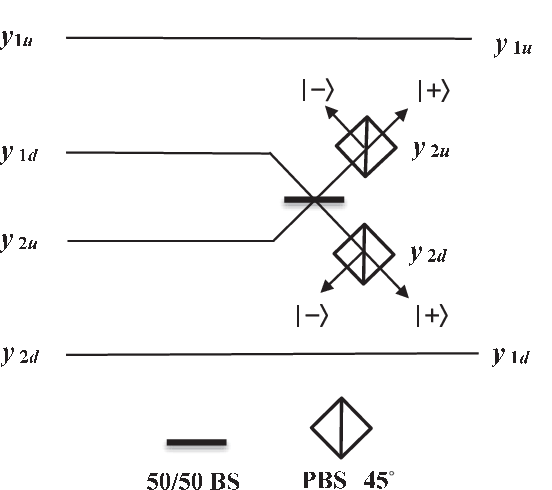}
\caption{The schematic diagram of the spatial mode parity check device. The photon $Y_1$ ($Y_2$) enters the apparatus via the two potential spatial modes $y_{1u}$ and $y_{1d}$ ($y_{2u}$ and $y_{2d}$). Photons in $y_{1d}$ and $y_{2u}$ interfere in the 50:50 beam splitter, after which the photons are measured in the polarization diagonal basis via the PBS at $45^o$. The name of $y_{2d}$ is changed to $y_{1d}$ for the selection purpose. }
\end{figure}
\end{center}

The schematic for the spatial mode parity check (SPC) is shown in Fig.2. It is used to discriminate the even parity states ($\vert uu\rangle$ or $\vert dd\rangle$) from the odd-parity ones ($\vert ud\rangle$ or $\vert du\rangle$). It is not difficult to verify that when two photons have different spatial modes, they will both appear at the output port 1 or 2. If they have the same spatial modes, each output port 1 and 2 has one and only one photon. 
A photon emitting from output port 2 is measured in the diagonal basis of the two DOFs.

With these two parity check devices, we can concentrate partially hyperentangled $N$-photon GHZ states that are entangled in the polarization and the spatial modes simultaneously.

\section{Hyperconcentration}

The partially hyperentangled $N$-photon GHZ state can be written as
\begin{eqnarray}
\vert \Psi\rangle_{AB...C}&=&(\alpha \vert HH...H\rangle +\beta\vert VV...V\rangle)\nonumber\\
&\otimes&(\delta\vert a_{u}b_{u}...c_{u}\rangle+\eta \vert a_{d}b_{d}...c_{d}\rangle).
\end{eqnarray}
The parameters satisfy the normalization conditions $\vert \alpha\vert ^2+\vert \beta\vert ^2=1$ and $\vert \delta\vert^2+\vert \eta\vert^2=1$. The subscripts $A$, $B$,...,$C$ represent the photons belonging to Alice, Bob,...,Charlie and $x_u$ and $x_d$ are the two potential spatial modes of photon $X (X=A,B,...C)$.
These $N$ parties can be spatially far apart.
The purpose of the hyperconcentration scheme is to obtain a maximally entangled state in both DOFs, i.e., the maximally hyperentangled GHZ state
\begin{eqnarray}
\vert \Phi\rangle_{AB...C}&=&\frac{1}{\sqrt{2}}(\vert HH...H\rangle +\vert VV...V\rangle)\nonumber\\&\otimes&\frac{1}{\sqrt{2}}(\vert a_{u}b_{u}...c_{u}\rangle+\vert a_{d}b_{d}...c_{d}\rangle).
\end{eqnarray}
To achieve this, we use two identical less-entangled
states to distill maximal hyperentanglement probabilistically. We start with two states $\vert \Psi\rangle_{A_1B_1...C_1}$ and $\vert \Psi\rangle_{A_2B_2...C_2}$.

First, we convert the second state to
\begin{eqnarray}
\vert \Psi\rangle_{A_2B_2...C_2}&=&(\alpha \vert VV...V\rangle +\beta\vert HH...H\rangle)\nonumber\\&\otimes&(\delta\vert a_{2d}b_{2d}...c_{2d}\rangle+\eta \vert a_{2u}b_{2u}...c_{2u}\rangle).
\end{eqnarray}
The flip of polarization state can be realized by half wave plates oriented at $45^o$. And the flip of spatial mode can be simply realized by changing their names.
The initial state of the $2N$-photon system can be written as
\begin{eqnarray}
&&\vert \Xi\rangle_{A_1B_1...C_1A_2B_2...C_2}\nonumber\\&=&[\alpha^2\vert HH...HVV...V\rangle+\beta^2\vert VV...VHH...H\rangle\nonumber\\&&+\alpha\beta(\vert HH...HHH...H \rangle+\vert VV...VVV...V\rangle)]\nonumber\\&\otimes& [\delta^2\vert a_{1u}b_{1u}...c_{1u}a_{2d}b_{2d}...c_{2d}\rangle
+\eta^2\vert a_{1d}b_{1d}...c_{1d}a_{2u}b_{2u}...c_{2u}\rangle\nonumber\\&&+\delta\eta(\vert a_{1u}b_{1u}...c_{1u}a_{2u}b_{2u}...c_{2u}\rangle+\vert a_{1d}b_{1d}...c_{1d}a_{2d}b_{2d}...c_{2d}\rangle)].
\end{eqnarray}
\begin{center}
\begin{figure}[!h]
\includegraphics[height=2in]{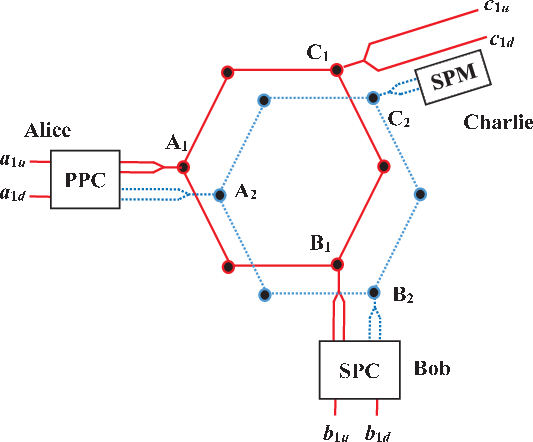}
\caption{Schematic diagram of the proposed hyperconcentration protocol. Maximal hyperentanglement is distilled from two identical less-entangled $N$-photon GHZ states (solid and dotted) distributed among $N$ parties who are spatially separated. Each party has a photon from each of the two less-entangled $N$-photon states.
One party (Alice) applies a PPC to check the polarization parity of $A_1$ and $A_2$ and another party (Bob) uses SPC to check the spatial mode parity of $B_1$ and $B_2$. For the other $N-2$ parties (Charlie, etc), single-photon two-qubit measurements (SPM) can be performed on their second photons.  For certain measurement results of $A_2$ and $B_2$, the quantum system collapses to the maximally hyperentangled GHZ state.}
\end{figure}
\end{center}
The schematic for hyperconcentration is shown in Fig.3. Firstly, one party, say Alice puts her qubits $A_1$ and $A_2$ into the polarization parity check device while the second party Bob guides his qubit $B_1$ and $B_2$ into the spatial mode parity check device. Both of them select the even-parity terms by requiring that both the two output ports 1 and 2 have one and exactly one photon. Then the selected state can be written as
\begin{eqnarray}
&&\vert \Xi'\rangle_{A_1B_1...C_1A_2B_2...C_2}\nonumber\\&=&\alpha\beta\delta\eta(\vert HH...HHH...H \rangle+\vert VV...VVV...V\rangle)\nonumber\\&\otimes& (\vert a_{1u}b_{1u}...c_{1u}a_{2u}b_{2u}...c_{2u}\rangle+\vert a_{1d}b_{1d}...c_{1d}a_{2d}b_{2d}...c_{2d}\rangle).
\end{eqnarray}
The probability of getting this state is $4\vert \alpha\beta\delta\eta\vert ^2$. After Alice and Bob's measurement on particle $A_2$ and $B_2$, the collapsed $(2N-2)$-photon state can be written as
\begin{eqnarray}
&&\vert \Xi''\rangle_{A_1B_1...C_1...C_2}\nonumber\\&=&\frac{1}{2}(\vert HH...H \rangle+(-1)^p\vert VV...V\rangle)\nonumber\\&\otimes& (\vert a_{1u}b_{1u}...c_{1u}...c_{2u}\rangle+(-1)^q\vert a_{1d}b_{1d}...c_{1d}...c_{2d}\rangle)
\end{eqnarray}
Here the parameters $p$ and $q$ depend on Alice and Bob's measurement results. If the polarization measurement results are $\vert +\rangle_{A_2}\vert +\rangle_{B_2}$ or $\vert -\rangle_{A_2}\vert -\rangle_{B_2}$, $p=0$. Otherwise, $p=1$. And if the spatial mode measurement results are both $d$ or $u$, $q=0$. Otherwise, $q=1$. To get the desired maximally hyperentangled $N$-photon state, each of the other parties performs a single-photon two-qubit measurement (SPM) on his/her second photon. The SPM setup is shown in Fig.4.
\begin{center}
\begin{figure}[!h]
\includegraphics[height=1.6in]{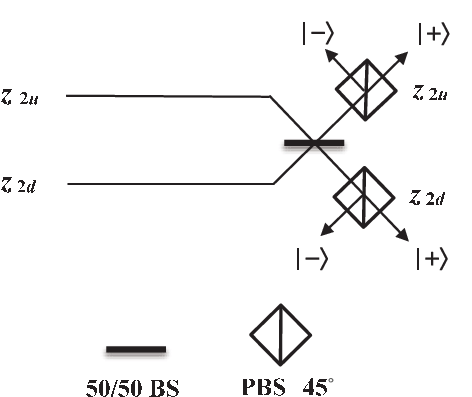}
\caption{Schematic diagram of the single-photon two-qubit measurement. Both the two DOFs of the single photon are measured in the diagonal basis. The measurement of the spatial mode is realized by the BS and of the polarization state by the PBS at $45^o$. }
\end{figure}
\end{center}

Then the final state is
\begin{eqnarray}
&&\vert \Xi'''\rangle_{A_1B_1...C_1}\nonumber\\&=&\frac{1}{2}(\vert HH...H \rangle+(-1)^P\vert VV...V\rangle)\nonumber\\&\otimes& (\vert a_{1u}b_{1u}...c_{1u}\rangle+(-1)^Q\vert a_{1d}b_{1d}...c_{1d}\rangle)
\end{eqnarray}
Here $P$ and $Q$ depend on all the $N$ parties' measurement outcomes. If the number of $\vert -\rangle$ is even (odd), $P=0$ $(1)$. And when the number of $d$ is even (odd), $Q=0$ $(1)$. Then one party can perform the phase-flip operation $\sigma_z=\vert H\rangle\langle H\vert-\vert V \rangle\langle V\vert$ ($\sigma_z=\vert x_1\rangle\langle x_1\vert-\vert x_2 \rangle\langle x_2\vert$) when $P=1$ $(Q=1)$ to obtain the desired state $\vert \Phi\rangle_{A_1B_1C_1}$. Thus, the $N$ parties can share the maximally hyperentangled state with a total success probability of  $4\vert \alpha\beta\delta\eta\vert ^2$, which is the same as the hyperconcentration scheme for the two-photon state \cite{hc1}. The relation between the success probability and the parameters of the initial states is shown in Fig.5.

\begin{center}
\begin{figure}[!h]
\includegraphics[height=2.5in]{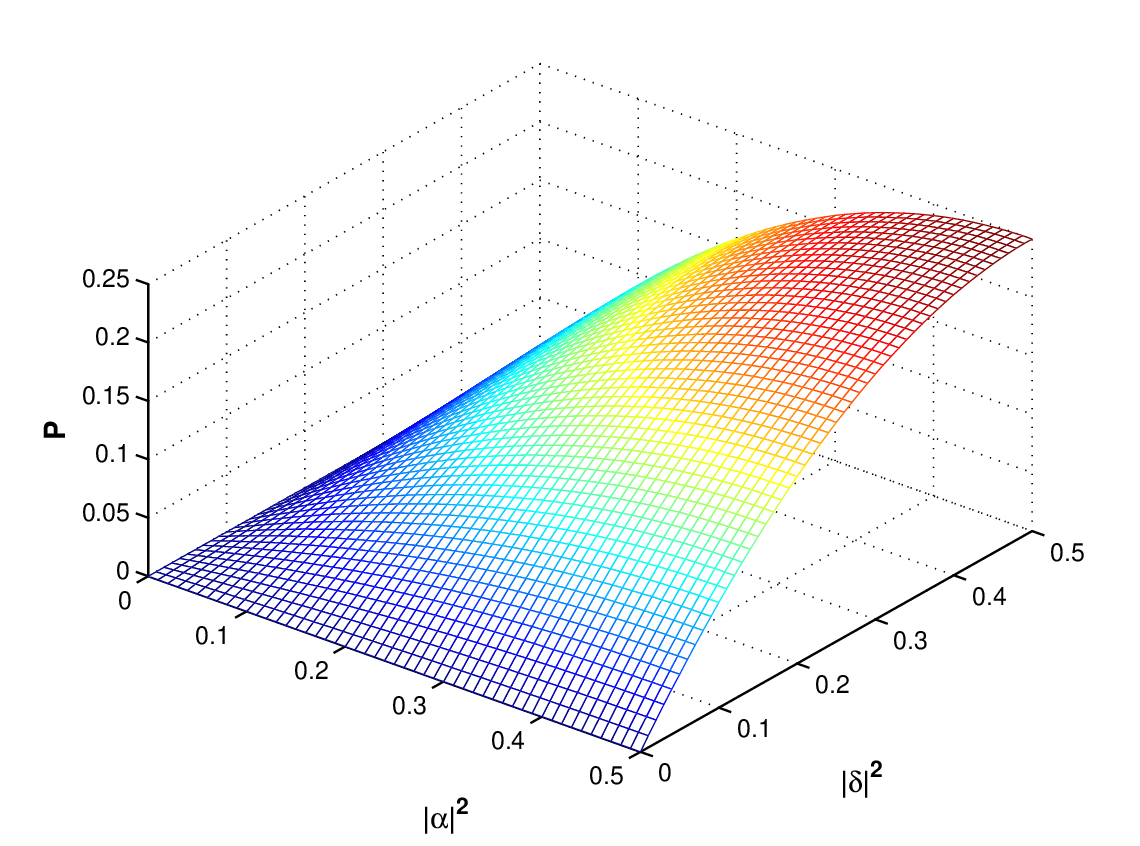}
\caption{The success probability of the hyperconcentration scheme as a function of the coefficients of the initial state.  }
\end{figure}
\end{center}

\section{Discussion and Summary}
We have proposed a hyperconcentration scheme for $N$-photon partially hyperentangled GHZ states. Each of the $N$ parties holds two photons, which come from two identical less-entangled states. One party performs the polarization parity check measurement while another performs the spatial mode parity check measurement. The other $N-2$ parties are only required to perform single-photon two-qubit measurements in the diagonal basis on both the two DOFs. By selecting the even-parity states in both the two parity check measurements, the remote $N$ parties can share the maximally hyperentangled GHZ state with a certain probability. The total success probability depends on the parameters of the initial states, with the maximum success probability being 25\%. The success probability is not optimal. This is due to the fact that maximal entanglement in both DOFs should be achieved. Since we restricted ourselves to linear optics, parity checks are based on the measurement of photons, after which, the photons are destroyed.  If nonlinear interactions which can realize quantum nondemolition (QND) measurements are utilized, failed instances can be reused iteratively and the success probability of hyperconcentration schemes can be improved \cite{hc11,hc2}. Fortunately, the success probability in our scheme does not decrease with the growth of photon number. The success probability  for the $N$-photon state is the same as that of the two-photon scheme \cite{hc1}.

There is another method for entanglement concentration called as the parameter splitting method \cite{hc1}, which also only requires linear optical elements. The common point of this method and ours is that both of them are practical and can be realized with current technology. Compared with our scheme, the parameter splitting method has a higher success probability. However, it requires the parameter information which is unknown in our protocol. The information can be obtained by measuring some samples. Therefore, which method is a better choice depends on the amount of states to be concentrated.   

In our scheme, only two parties perform the parity check. However, it can be changed to $N$ parity checks, among which at least one PPC and one SPC are required. The success probability is the same since after the success of the first PPC and the first SPC the remaining parity check measurements will definitely give the even-parity results.

The success of our hyperconcentration scheme is based on the two parity check measurements. In principle, one and only one detector clicks in each parity check measurement, which is the signal of success. However, according to Eq.(5), there exists the probability that two photons arrive at the same detector, which may causes an error if the detectors cannot distinguish between one and two photons. One solution is using the photon number resolving (PNR) detector. However, it is expensive and uncommon.
Another solution to eliminate this error is using an improved parity check devices. For the polarization parity check, two PBSs are introduced so that two photons with different polarizations will trigger two separate detectors. The improved PPC is shown in Fig.6. For the spatial mode parity check, additional BSs before the PBS at $45^o$ can reduce the probability of errors. Imperfect optical elements and detectors will nevertheless decrease the success probability and cause errors in practice. In this case, postselection is required to solve these problems by strictly selecting the case in which each output port 1 and 2 has one and only one photon.

\begin{center}
\begin{figure}[!h]
\includegraphics[height=2.2in]{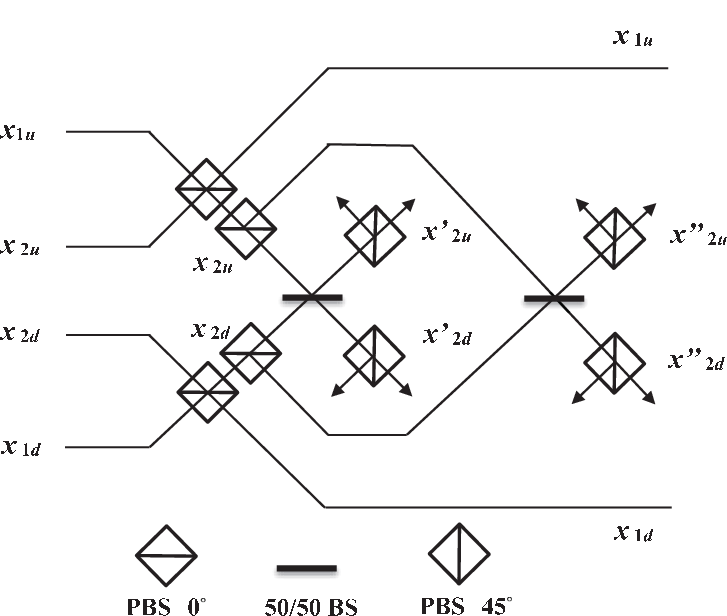}
\caption{Schematic diagram of the improved polarization parity check device. Two PBSs at $0^o$ were added to avoid the situation that two photons go to the same detector. }
\end{figure}
\end{center}

In our scheme, since the parameters are unknown, two less-entangled states are employed to distill one maximal hyperentangled GHZ state probabilistically. When $N$ is large, this method consumes $2N$ photons to obtain an $N$-photon state probabilistically. A more practical approach would be to obtain information about the unknown parameters by measuring a sufficient number of sample states. Then given the known parameters, an additional quantum state can be produced to assist the concentration. The auxiliary state required is
\begin{eqnarray}
\vert \varphi\rangle_{A_2B_2}=(\alpha\vert VV\rangle+\beta \vert HH\rangle)\otimes(\delta\vert a_{2d}b_{2d}\rangle+\eta\vert a_{2u}b_{2u}\rangle)
\end{eqnarray}
The principle is shown is Fig.7. Alice and Bob perform PPC and SPC on $A_1, A_2$ and $B_1, B_2$, respectively. By keeping the even-parity state in both these two DOFs, the $N$ parties share the maximal hyperentangled state. The success probability is also $4\vert \alpha\beta\delta\eta\vert ^2$. However, this method requires that the two parties Alice and Bob be located at the same place.
\begin{center}
\begin{figure}[!h]
\includegraphics[height=2in]{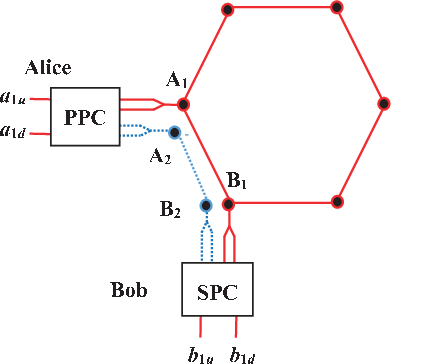}
\caption{Schematic diagram of the hyperconcentration protocol assisted by two auxiliary photons. Alice performs a polarization parity check on $A_1$ and $A_2$ while SPC is used to check the spatial mode parity of $B_1$ and $B_2$. By selecting on the even-parity states in both these two DOFs, the quantum system collapses to the maximally hyperentangled GHZ state.}
\end{figure}
\end{center}

To sum up, we have presented a hyperconcentration scheme for an $N$-photon hyperentangled state based on one polarization parity check measurement and one spatial mode parity check measurement, which only require linear optical elements. And for the remaining $N-2$ parities, only single-qubit measurements are required. This method can help remote parties share the maximally hyperentangled GHZ state which can be used in subsequent quantum information processing. The success probability is the same as that of  the two-photon hyperconcentration scheme. These features can make our protocol more useful for practical applications in long-distance quantum communication.

\section*{Acknowledgement}
XL is supported by the National Natural Science
Foundation of China under Grant No. 11004258 and the Fundamental Research Funds for the Central Universities under
Grant No.CQDXWL-2012-014. SG acknowledges support from the Ontario Ministry of Research and Innovation and the Natural Sciences and Engineering Research Council of Canada.

\end{document}